\documentclass[twocolumn]{aastex631}

\usepackage{amsmath}
\allowdisplaybreaks[1]

\usepackage{color}
\newcommand\denyz[1]{\textcolor{magenta}{[#1]}}

\usepackage{multirow}

\shorttitle{2BRlx+TDE}
\shortauthors{Melchor et al.}
\graphicspath{{./}{figures/}}

\begin{document}

\title{Tidal Disruption Events from the Combined Effects of Two-Body Relaxation and the Eccentric Kozai-Lidov Mechanism}


\correspondingauthor{Denyz Melchor}
\email{denyzamelchor@astro.ucla.edu}

\author[0000-0002-7854-1953]{Denyz Melchor}
\affiliation{Department of Physics and Astronomy, University of California, Los Angeles, CA 90095, USA}
\affiliation{Mani L. Bhaumik Institute for Theoretical Physics, Department of Physics and Astronomy, UCLA, Los Angeles, CA 90095, USA}

\author[0000-0001-6350-8168]{Brenna Mockler}
\affiliation{The Observatories of the Carnegie Institution for Science, Pasadena, CA 91101, USA}
\affiliation{Department of Physics and Astronomy, University of California, Los Angeles, CA 90095, USA}
\author[0000-0002-9802-9279]{Smadar Naoz}
\affiliation{Department of Physics and Astronomy, University of California, Los Angeles, CA 90095, USA}
\affiliation{Mani L. Bhaumik Institute for Theoretical Physics, Department of Physics and Astronomy, UCLA, Los Angeles, CA 90095, USA}
\author[0000-0003-0984-4456]{Sanaea Rose}
\affiliation{Department of Physics and Astronomy, University of California, Los Angeles, CA 90095, USA}
\affiliation{Mani L. Bhaumik Institute for Theoretical Physics, Department of Physics and Astronomy, UCLA, Los Angeles, CA 90095, USA}
\author[0000-0003-2558-3102]{Enrico Ramirez-Ruiz}
\affiliation{Department of Astronomy and Astrophysics, University of California, Santa Cruz, CA 95064, USA}

\begin{abstract}
Tidal disruption events (TDEs) take place when a star ventures too close to a supermassive black hole (SMBH) and becomes ruptured. One of the leading proposed physical mechanisms often invoked in the literature involves weak two-body interactions experienced by the population of stars within the host SMBH's sphere of influence, commonly referred to as two-body relaxation. This process can alter the angular momentum of stars at large distances and place them into nearly radial orbits, thus driving them to disruption. On the other hand, gravitational perturbations from an SMBH companion via the eccentric Kozai-Lidov (EKL) mechanism have also been proposed as a promising stellar disruption channel. Here we demonstrate that the combination of EKL and two-body relaxation in SMBH binaries is imperative for building a comprehensive picture of the rates of TDEs. Here we explore  how the density profile of the surrounding stellar distribution and the binary orbital parameters of an SMBH companion influence the rate of TDEs. We show that this combined channel naturally  produces disruptions at a rate that is consistent with observations and also naturally forms repeated TDEs, where a bound star is partially disrupted over multiple orbits. Recent observations show stars being disrupted in short-period  orbits, which is challenging to explain when these mechanisms are considered independently. However, the diffusive effect of two-body relaxation, combined with the secular nature of the eccentricity excitations from EKL, is found to drive stars on short eccentric orbits at a much higher rate. 

\end{abstract}


\section{Introduction} \label{sec:intro}

Tidal Disruption Events (TDEs) occur when a star passes near a supermassive black hole (SMBH) and gets torn apart by tidal forces \citep[e.g.,][]{hills_possible_1975,rees_tidal_1988,2013ApJ...767...25G}. As the star begins to be torn apart, a fraction of it may form 
an accretion disk which may result in an electromagnetic signature \citep[][]{rees_tidal_1988,evans_tidal_1989,ulmer_flares_1999,2014ApJ...783...23G}.
Thus, TDEs are promising signatures for understanding  stellar population around SMBHs as well as accretion processes onto SMBHs \citep[e.g.,][]{dai_physics_2021,gezari_tidal_2021,2022ApJ...924...70M}. The rate of these events seems also to have many repercussions for SMBH and host galaxy demographics \citep[e.g.][]{van_velzen_measurement_2014,kochanek_tidal_2016,gezari_tidal_2021,2017ApJ...850...22L,2020SSRv..216...32F,2021ApJ...907L..21D}. For instance it has been suggested that the TDE rate  can be used to discriminate between SMBH formation scenarios \citep[][]{stone_rates_2016} and help probe the spin distribution of massive SMBHs \citep[][]{kesden_tidal_2012,2020ApJ...905..141L}. Despite these exciting prospects, many challenges exist in effectively estimating TDE rates. 



Previous studies calculating rates of TDEs have often focused on the two-body relaxation process. In this process, weak gravitational interactions, {\it kicks} from neighboring stars over long periods of time are able to place stars in near radial  orbits  around an SMBH \citep[][]{frank_effects_1976,rees_tidal_1988,Rauch+96}. This channel receives a lot of attention, both analytically and numerically \citep[][]{magorrian_rates_1999, wang_revised_2004,brockamp_tidal_2011,2012ApJ...757..134M,stone_rates_2016}. Notably, these studies estimate the rate at which  stars can be positioned  in orbit with pericenter distances comparable to or smaller than their corresponding  tidal radius \citep{frank_effects_1976}. The estimated rates, both numerical and analytical, are in agreement that about $10^{-5} - 10^{-4}$ stars per year will undergo a TDE in a typical galaxy.



 Another channel often considered in the literature involves the presence of an SMBH binary. The hierarchical nature of galaxy formation, combined with SMBHs residing in the centers of almost every galaxy, implies that SMBH binaries are prevalent in our Universe \citep[e.g.,][]{Begelman+80,di_matteo_energy_2005,Hopkins+06,Robertson+06,Callegari+09,li_pairing_2020}. Furthermore, observations of dual active galactic nuclei (AGN), typically a few kiloparsecs or more apart, seem to suggest that these configurations will eventually lead to a tight SMBH binary \citep[e.g.,][]{Komossa+03,Bianchi+08,Comerford+09bin,Comerford+18,Green+10,Smith+10,foord_second_2020,Stemo+20,stemo_catalog_2021,Liu+10kpc,Li+20Pair}.
Closer to home, theoretical arguments combined with observational campaigns insinuate that our galactic center may also host a massive black hole companion \citep[e.g.,][]{Hansen+03,Maillard+04,gurkan_disruption_2005,gualandris_perturbations_2009,Chen+13,Fragione+20,Gravity+20,Generozov+20,Naoz+20,Zheng+20}.

Gravitational perturbations from an SMBH companion can significantly modify the orbits of surrounding stars \citep[e.g.,][]{chen_tidal_2008,chen_enhanced_2009,Chen+11,Chen+13}. Most notably, the  EKL mechanism \citep{Kozai,Lidov,Naoz16} has been shown to excite the eccentricities of stars to high values \citep[][]{Li+14,li_eccentricity_2014,li_implications_2015}. The EKL-channel is expected to result in a burst-like TDE rate, \citep{Mockler+22}, of tens to hundreds of times higher than  two-body relaxation alone for $\approx 10^7$ years.

Recent observations of {\it repeating} tidal disruption events (rTDE) have raised the question of why none of the aforementioned channels, two-body relaxation or EKL, predicted  a sizable number of rTDEs. In an rTDE, a star is partially disrupted and may experience multiple disruptive events \citep[][]{2013ApJ...777..133M,2014ApJ...794....9M,campana_multiple_2015,payne_asassn-14ko_2021,payne_rapid_2022}. A popular example is ASASSN-14ko, which is a periodically flaring transient, every $\approx 115 $ days \cite[][]{payne_asassn-14ko_2021,2023ApJ...944..184L}. How do these stars migrate to such close distances around the SMBH without becoming fully disrupted?
The two-body relaxation process drives stars on large separations from the SMBH onto nearly radial orbits \citep[e.g.,][]{Fragione+18,Sari+19}; therefore, these stars have large semi-major axes and low angular momentum. The EKL mechanism changes the angular momentum of an orbit and can drive the eccentricity to extreme values \citep[e.g.,][]{Naoz+11sec} while keeping the star separation constant. At face value, both of these channels seem to face challenges in explaining rTDEs. Specifically, for repeated TDEs, the star separation needs to be small, $10^{-3}$ to $10^{-2}$~pc for SMBH with masses in the range of $10^{6.5-7.86}$~M$\odot$ in order to have an orbital period ranging from $115$~days to $30$~years \citep[e.g.,][]{payne_asassn-14ko_2021,liu_deciphering_2023,malyali_rebrightening_2023,wevers_live_2023}. Another possible formation channel for rTDEs is the widely discussed  binary disruption and capture via the Hills mechanism. Interestingly, eccentric stellar disks can yield promising breeding grounds for a high rate of binary disruptions. Having said this,  the disruption of highly bound binaries is required in order to explain a bound star  with an orbital period $\approx 115$~days.  




Following \citet{naoz_combined_2022}, we propose a novel mechanism combining gravitational perturbations from a far-away SMBH companion with weak two-body interactions from the overall population of stars around the primary SMBH. The combination of these two mechanisms, EKL, and two-body relaxation, is necessary for the formation of rTDEs. While EKL, on its own, conserves the energy of any individual orbit, two-body relaxation describes the diffusive changes of the star's energy and angular momentum. Therefore,  diffusive effects  can drive a star's orbit to a part of the parameter space that is highly sensitive to EKL, thus triggering eccentricity and inclination excitations, which can more easily result in  a rTDE. Here we show that this combined mechanism not only creates a more continuous TDE rate, in contrast to the burst-like EKL channel found in \citet{Mockler+22}, but also naturally forms a large relative number of rTDEs. The paper is organized as follows: We describe the basic physical concepts in Section \ref{sec:Methods}. Then, we present our numerical simulations, both for a representative system and then for a large number of runs  in Section \ref{sec:Sims}. In Section \ref{sec:predictions}, we describe the TDE rate and formation of rTDEs using this novel mechanism. Finally, we end with a discussion of our findings in Section \ref{sec:discussion}.

\section{Basic Concepts and Characteristic Timescales}\label{sec:Methods}
\begin{figure}
  \begin{center} 
    \includegraphics[width=\linewidth]{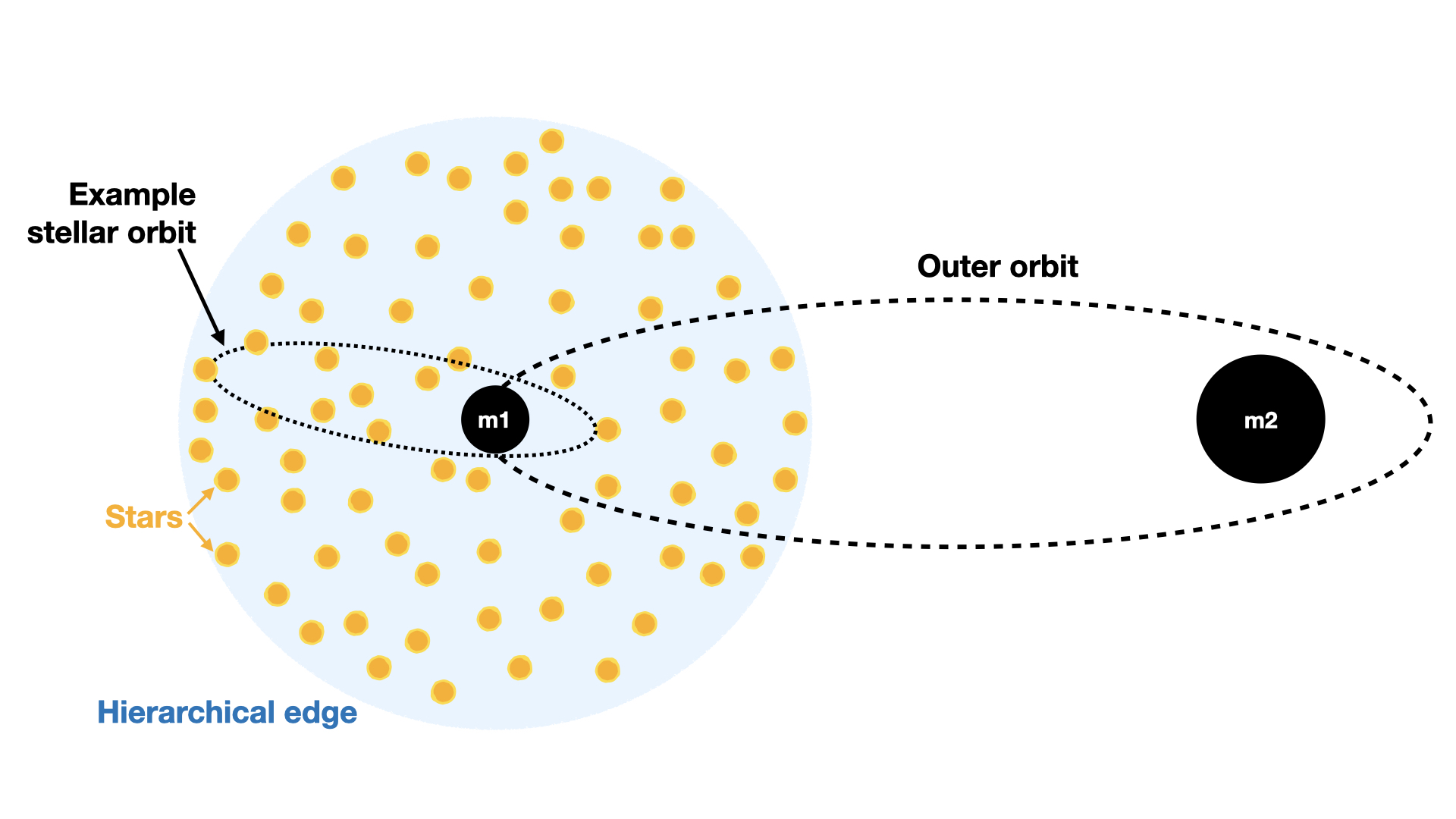}
  \end{center} 
  \caption{  \upshape {\bf The hierarchical triple system.}  {This hierarchical triple includes an SMBH binary of masses $m_1$ and $m_2$ on an outer orbit, and stellar cluster with components $m_\star$ on an inner orbit around $m_1$.}  } \label{fig:system} 
\end{figure}

We consider an SMBH with mass $m_1$ and an SMBH companion $m_2$ with a  semi-major axis $a_{\rm bin}$ and eccentricity $e_{\rm bin}$. In this system, we select $m_1 < m_2$, where the primary SMBH was fixed to mass $m_1 = 10^7$~M$_{\odot}$, while the secondary SMBH mass was varied between $m_2$ = 10$^{8}$~M$_{\odot}$ and $10^9$~M$_{\odot}$. 
Surrounding $m_1$ is a population of stars with mass $m_\star \approx 0.8$~M$_{\odot}$, each, and separation $r_\star=a_\star (1-e_\star^2) / (1+e_\star \cos f_\star)$, where $a_\star$, $e_\star$ and $f_\star$ are the star's semi-major axis, eccentricity, and true anomaly, respectively.  The density profile $\rho_\star(r)$ of the stellar components is calibrated by the M-$\sigma$ relation \cite[e.g.][]{Tremaine+02}:
\begin{equation}
    \label{eq:rhostar}
    \rho_{\star}(r)= \frac{ m_1}{\langle m_\star \rangle}\left(\frac{G\sqrt{m_1 M_0}}{\sigma_0^2 r}\right)^{-3+\alpha} \nonumber \ ,
\end{equation}
where $M_0=10^8$~M$_\odot$ and $\sigma_0=200$~km~sec$^{-1}$, are scaling factors. We study two nominal density profiles, $\alpha=1$ and $\alpha=2$, corresponding to a core and cusp stellar distribution, respectively. The maximum separation of the stars is defined by the hierarchical edge set by
\citep[e.g.,][]{LN11}:
 \begin{equation}\label{eq:epsilon}
     \epsilon = \frac{a_\star}{a_{\rm bin}}\frac{e_{\rm bin}}{1-e_{\rm bin}^2} \ ,
 \end{equation}
 as depicted in Figure \ref{fig:system}.

\begin{figure}
  \begin{center} 
    \includegraphics[width=\linewidth]{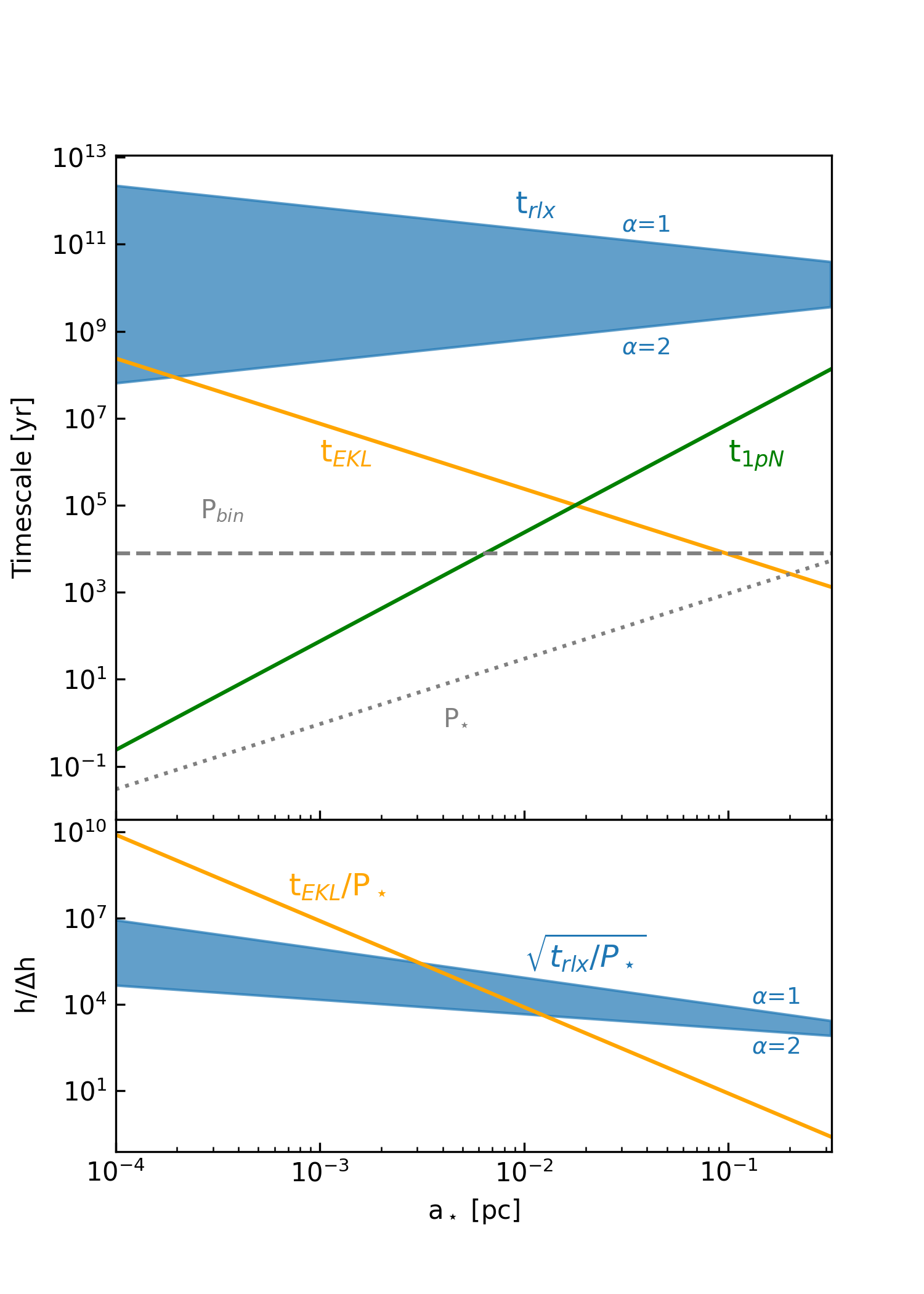}
  \end{center} 
  \caption{  \upshape {\bf Comparing characteristic  timescales.}  
  {\textit{Top panel:} In orange we show the EKL timescale for an $m_1$ = 10$^7$~M$_\odot$ and $m_2$ = 10$^9$~M$_\odot$ SMBH binary and a population of stars of $m_\star$ = 0.8~M$_\odot$. In blue we show the relaxation timescale for the inner binary of $m_1$ and $m_\star$, allowing $\alpha$ to range between 1 and 2. The dotted grey line is the star's orbital period around $m_1$, while the dashed line is the period of the SMBH binary.
  \textit{Bottom panel:} Here we compare the relative change in angular momentum $h$. In orange $h/\Delta h \approx t_{EKL}/P_\star$ for the EKL mechanism and in blue $h/\Delta h \approx \sqrt{t_{rlx}/P_\star}$} for two-body relaxation. Further description is provided at the end of Section \ref{sec:Methods}. See \citet{Naoz+22} for a similar analysis. } \label{fig:timescale} 
\end{figure}

\begin{figure}
  \begin{center} 
    \includegraphics[width=\linewidth]{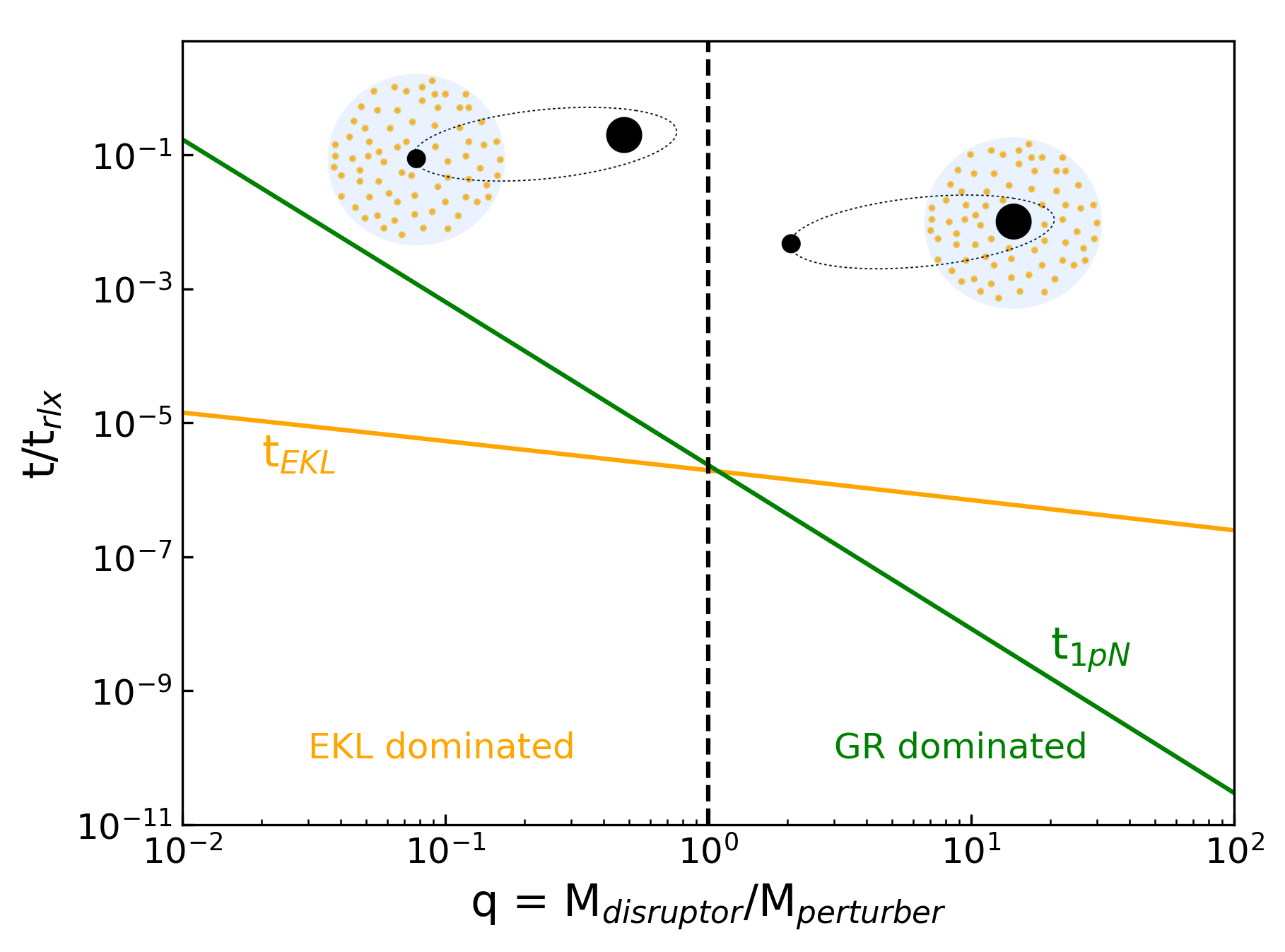}
  \end{center} 
  \caption{  \upshape {\bf Principal timescales in a SMBH binary.} Here, we normalize the timescales by the two-body relaxation timescale (Eq.~\ref{eq:trelx}) and depict the EKL and the 1pN timescales (Eq.~\ref{eq:tEKL} and Eq.~\ref{eq:t1PN}, respectively) as a function of the mass ratio $q$. To highlight the dependency on the mass ratio, we explicitly denote $m_{\rm disruptor}$ as the one that forms the TDE and $m_{\rm perturber}$ as the far away SMBH companion. To generate this figure, we adopted $m_{\rm perturber}=10^8$~M$_\odot$ and vary $m_{\rm disruptor}$ between $10^{6}$M$_\odot$ and $10^{10}$~M$_\odot$ with $e_{\rm bin} = 0.7$ and $e_{\star}$ = 0.9. The binary is set at $a_{bin} = 1$~pc and the star at $a_{\star} = 0.07$~pc. The dashed line denotes  $q = 1$.   To the right of the dashed line we have the regime where $m_{\rm disruptor} > m_{\rm perturber}$, and the 1pN precession suppresses the EKL eccentricity excitations. For $m_{\rm disruptor} < m_{\rm perturber}$, on the other hand, EKL excitations are dominant. In the presence of two-body relaxation, high eccentricities can be excited to larger values (Figure \ref{fig:timescale}). Throughout the paper, we thus focus on the left regime, which can excite eccentricities more efficiently. 
  } \label{fig:regimes} 
\end{figure}

We solve the secular, hierarchical three-body equations up to the octupole level of approximation following \citet{Naoz+11sec}. The EKL mechanism can excite the stars' eccentricity and inclination as a time function. These eccentricity excitations can lead to TDEs if the stars' pericenter $a_{\star}(1-e_{\star})$, crosses the tidal threshold 
\begin{equation}\label{eq:Rcrit}
    R_{\rm T} \approx R_\star \left(\frac{m_1}{m_\star}\right)^{1/3} \ ,
\end{equation}
where $R_\star$ is the radius of the star. Here we assume a population of stars with $0.8$~M$_\odot$, which  correspond to $R_\star=0.7~$R$_\odot$.  As mentioned, previous studies showed that the effect of EKL is efficient in creating TDEs for stars distributed around the less massive SMBH \citep[e.g.,][]{Li+15,Mockler+22}. The relevant timescale for these events is \citep[e.g.,][]{Antognini15} estimated by:
\begin{equation}\label{eq:tEKL}
    t_{\rm EKL} \approx \frac{16}{30\pi} \frac{m_1+m_\star + m_2}{m_2}\frac{P_{\rm bin}^2}{P_\star}(1-e_{\rm bin}^2)^{3/2} \ .
\end{equation}
This timescale is shown in Figure \ref{fig:timescale}. 

It has been suggested that the EKL mechanism is less efficient in driving high eccentricity of the stars around the more massive SMBH because general relativistic (GR) precession can suppress the eccentricity excitation \citep[e.g.][]{Ford00Pls,Naoz+12GR,Li+15,Mockler+22}. The 1st post-Newtonian precession takes place on the following timescale:
\begin{equation}\label{eq:t1PN}
  t_{\rm 1pN} \approx \frac{P_\star c^2 a_\star(1-e_\star^2)}{6\pi G (m_1+m_\star)}
  \end{equation}
where $c$ is the speed of light. The relevant timescale is shown in Figure \ref{fig:timescale}. 
Large eccentricity excitations will take place when the EKL timescale is shorter than the GR precession timescale. 
We illustrate this in Figure \ref{fig:regimes}, which depicts the  EKL and GR precession timescales normalized to the relaxation timescale as a function of the mass ratio. We define the mass ratio as $q=m_{\rm disruptor}/m_{\rm perturber}$. As shown in the Figure, for $m_{\rm disruptor} > m_{\rm perturber}$, GR precession dominates the EKL eccentricity excitations, thus suppressing TDE formation. On the other hand, when  $m_{\rm disruptor} < m_{\rm perturber}$, the EKL eccentricity excitation timescale is faster than GR precession. This trend motivates the choice of having $m_1<m_2$ in our analysis. 

We note that the SMBH binary orbital timescale is comparable to the EKL timescale in some parts of the parameter space. At face value, this may suggest that the double average approach adopted here leads to misleading results. However, as was shown by \citet{Antonini+14,Antognini+14}, and \citet{Luo+16}, the main difference between our double average approach and an N-body approach (for these setups) 
is that N-body simulations can result in even higher eccentricity excitations. Thus, we estimate that the eccentricity values achieved below represent a lower limit.  

As mentioned before, two-body relaxation processes were proposed as a promising channel to produce TDEs. The typical timescale to change the orbit's angular momentum ($h$) and the orbital energy by an order of themselves is estimated by \citep[e.g.,][]{binney_galactic_2008}: 
\begin{equation}\label{eq:trelx}
    t_{\rm rlx} = 0.34 \frac{\sigma_\star^3}{G^2 \rho_\star\langle m_{\rm scat} \rangle\ln\Lambda} \ ,
\end{equation}
where $\langle m_{\rm scat}\rangle$ is the mass of the average star that acts as a scatterer, $\sigma_\star$ is the velocity dispersion of stars around the SMBH
\begin{equation}
    \sigma_\star^2=\frac{Gm_1}{r_\star(1+\alpha)} \ ,
\end{equation}
where $\alpha$ is the slope of the stellar density profile. Lastly, the coulomb logarithm is:
\begin{equation}
    \Lambda = \frac{r_\star \sigma_\star^2}{2G \langle m_{\rm scat}\rangle}  \ .
    \end{equation} 
We adopt $m_{scat} = m_\star=0.8$~M$_\odot$. We show this timescale (Eq.~(\ref{eq:trelx})), in Figure \ref{fig:timescale}, for different density profiles $\alpha$, between $1$ and $2$.

The weak two-body relaxation processes are often neglected in the literature because the relevant timescale is larger compared to the EKL timescale (see Figure \ref{fig:timescale}). However, as shown recently, comparing the timescales can be misleading, and instead, the change applied to the angular momentum should be compared \citep{Naoz+22}. As depicted in Figure \ref{fig:timescale}, bottom panel, when comparing the $h/\Delta h$, in a large part of the parameter space, the two-body relaxation for a cusp-like profile, $\alpha=2$, yields changes to the angular momentum $h/ \Delta h$ comparable, or even shorter than the EKL's.

Following \citet{Naoz+22},
we model the change to the stellar orbit's velocity 
$v_\star=\sqrt{Gm_1 ( 2/r_\star - 1/a_\star)}$ due to one encounter as a random walk with isotropically oriented kicks to the stellar velocity once per orbit around the SMBH. The kick is assumed to be instantaneous at some random phase of the orbit about the SMBH.  Each  component of this three-dimensional kick is drawn from a Gaussian distribution with a zero average and a standard deviation of $\Delta v/\sqrt{3}$, where
\begin{equation}\label{eq:Deltav}
    \Delta v = v_\star\sqrt{ \frac{P_\star}{t_{\rm relx}}} \ , 
\end{equation}
\citep[see also][]{Bradnick+17}. See \citet{Naoz+22} for the complete set of equations.

\begin{figure}
  \begin{center} 
    \includegraphics[width=\linewidth]{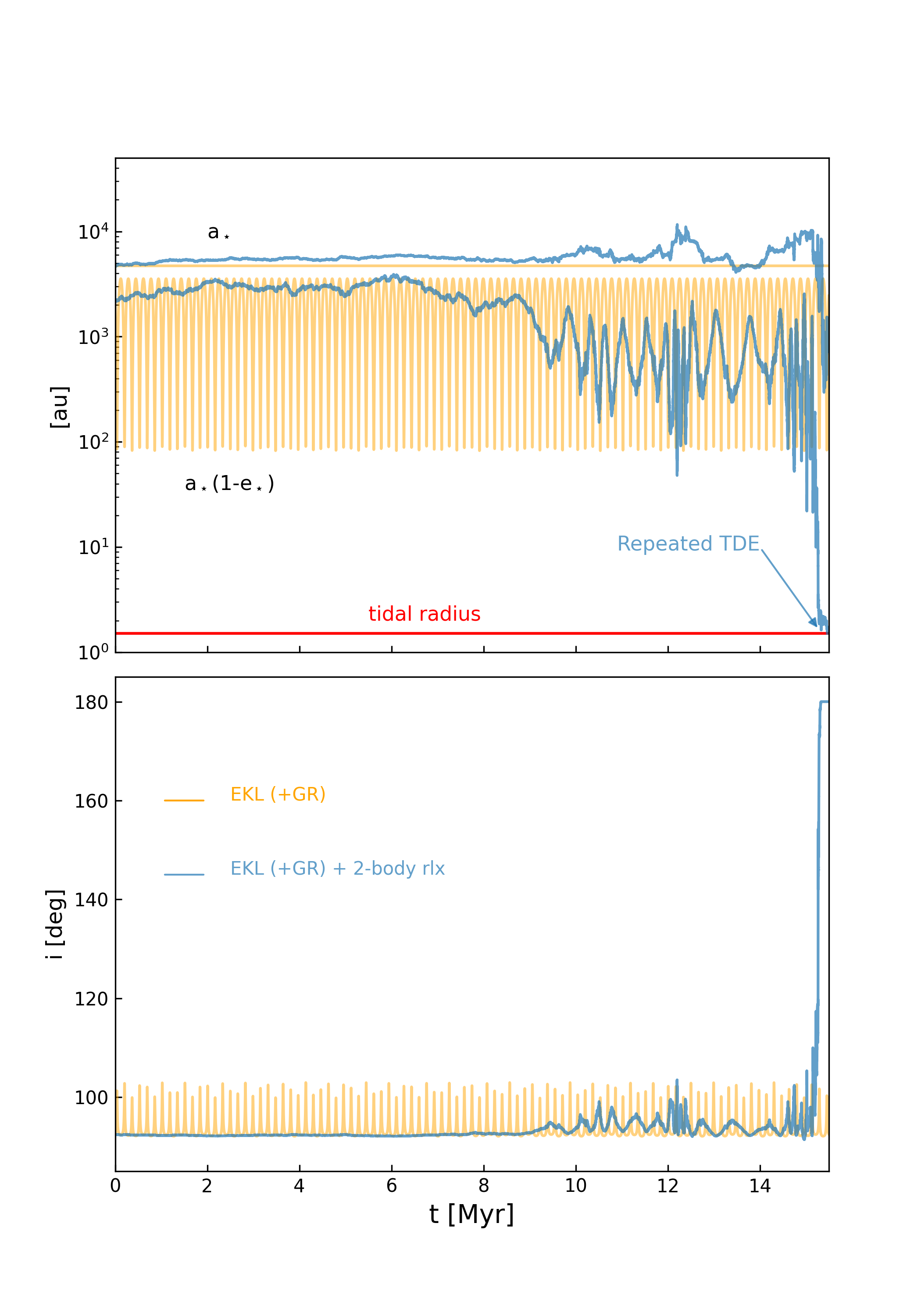}
  \end{center} 
  \caption{  \upshape {\bf An example of a repeated TDE. }
  {The time evolution of a supermassive black hole binary is shown. The primary black hole is of mass $m_1$ = 10$^{7}$~M$_\odot$ with a secondary of $m_2$ = 10$^9$~M$_\odot$. There is a stellar mass of $m_\star$ = 0.8~M$_\odot$ that is gravitationally bound to the primary at a separation of $4.7 \times 10^3$~au. The top panel compares the changes to the semi-major axis and pericenter radius of the inner binary when the system undergoes only EKL (+GR) (yellow) and EKL (+GR) with two-body relaxation (blue). The tidal radius is shown in red, where in crossing this line, we consider the system to have become a TDE. The semi-major axis is shown to drop considerably in the EKL and two-body relaxation simulation, indicating that this system could potentially be a repeated TDE.}  } \label{fig:timeEvolution} 
\end{figure}

\begin{table*}
\centering
\caption{Simulation parameters}
    \renewcommand{\thefootnote}{\arabic{footnote}}
    \footnotesize
    \setlength\tabcolsep{1.2pt}
    \renewcommand{\arraystretch}{1.5}
\begin{tabular}{cccccccc}
\hline
 
  run \#  & $q$  & $e_{\rm bin}$ &   \hspace{0.1cm} $\alpha$ & \hspace{0.1cm} $f_{\rm TDE}$ & \hspace{0.1cm} $f_{\rm TDE}$  $+$ & $f_{\rm rTDE}$ & \hspace{0.05cm} $N_{*}(\leq r_{\rm max})$ \\
   &   & & &  & $f_{\rm non-hierarchical}$ & &\\

\hline
0 & 10 & 0.3   &  \hspace{0.1cm} 1  & \hspace{0.1cm} 0.18 - 0.33 & 0.85 & - & 6e5  \\ 
1 & 10 & 0.3   &  \hspace{0.1cm} 2  & \hspace{0.1cm} 0.22 - 0.31 & 0.98 & 0.0156 & 4e6  \\
2 & 100 & 0.3   &  \hspace{0.1cm} 1  & \hspace{0.1cm} 0.035 - 0.37 & 0.82 & - & 6e5  \\ 
3 & 100 & 0.3   &  \hspace{0.1cm} 2  & \hspace{0.1cm} 0.15 - 0.47 & 0.99 & 0.0101 & 4e6 \\ 
4 & 10 & 0.5   &  \hspace{0.1cm} 1  & \hspace{0.1cm} 0.24 - 0.41 & 0.84 & - & 1e5  \\ 
5 & 10 & 0.5   &  \hspace{0.1cm} 2  & \hspace{0.1cm} 0.21 - 0.30 & 0.99 & 0.0260 & 2e6  \\ 
6 & 100 & 0.5   &  \hspace{0.1cm} 1  & \hspace{0.1cm} 0.056 - 0.51 & 0.87 & - & 1e5  \\ 
7 & 100 & 0.5   &  \hspace{0.1cm} 2  & \hspace{0.1cm} 0.18 - 0.48 & 1.0 & 0.0122 & 2e6  \\ 
8 & 10 & 0.7   &  \hspace{0.1cm} 1  & \hspace{0.1cm} 0.20 - 0.48 & 0.81 & - & 3e4   \\ 
9 & 10 & 0.7   &  \hspace{0.1cm} 2  & \hspace{0.1cm} 0.14 - 0.29 & 1.0 & 0.0238 & 9e5   \\ 
10 & 100 & 0.7  &  \hspace{0.1cm} 1  & \hspace{0.1cm} 0.034 - 0.62 & 0.87 & - & 3e4  \\ 
11 & 100 & 0.7  &  \hspace{0.1cm} 2  & \hspace{0.1cm} 0.087 - 0.50 & 1.0 & 0.0196 & 9e5    \\ 

\hline
\end{tabular}
\label{table:simparams}
\end{table*}



\section{Numerical Simulations }\label{sec:Sims}

\subsection{Dynamical Evolution of a Representative Example}\label{sec:OneSystem}
In Figure \ref{fig:timeEvolution}, we consider our fiducial model of $m_1=10^7$~M$_\odot$, and $m_2=10^9$~M$_\odot$, set on $a_{\rm bin}=4\times10^5$~au, and $e_{\rm bin}=0.5$. The star is initialized with $a_\star=4711$~au, $e_\star=0.53$, $\Omega=168^\circ$ and $\omega=38^\circ$, where $\Omega$ and $\omega$ are the longitude of ascending nodes and the argument of periapsis, respectively. For this example, we depict two cases, one which only includes EKL + GR, yellow, and EKL + GR + two-body relaxation processes, blue. We note that while we limit the EKL + GR presentation to $15$~Myr, to provide a comprehensive comparison, the system never had its eccentricity excited to cause a TDE. However, in the EKL + GR + two-body relaxation run, the star's pericenter crossed the tidal radius signifying a TDE. 

The final separation of $m_1$ and $m_\star$ decreases to $877$~au, crossing the tidal radius. At around $10^7$~years the system begins eccentricity oscillations which change the inclination and semi-major axis of the primary's orbit with the stellar mass. By $15$~Myr, the system has reached an eccentricity of $0.998$, resulting in a nearly $90^\circ$ rotation in the inclination. The stellar mass  can potentially become a repeated tidal disruption event by $m_1$ and continue to orbit the SMBH, where this rTDE would be seen every $P_\star = 9$~yr.

\subsection{Monte-Carlo simulations}\label{sec:MonteCarlo}

We ran a total of 12,000 runs for 12 realizations varying the SMBH binary eccentricity, mass ratio, and the power law of the stellar density profile. The system was initialized with $m_1 = 10^7$~M$_\odot$ and $m_\star = 0.8$~M$_\odot$.

\begin{figure*}
  \begin{center} 
    \includegraphics[width=\linewidth]{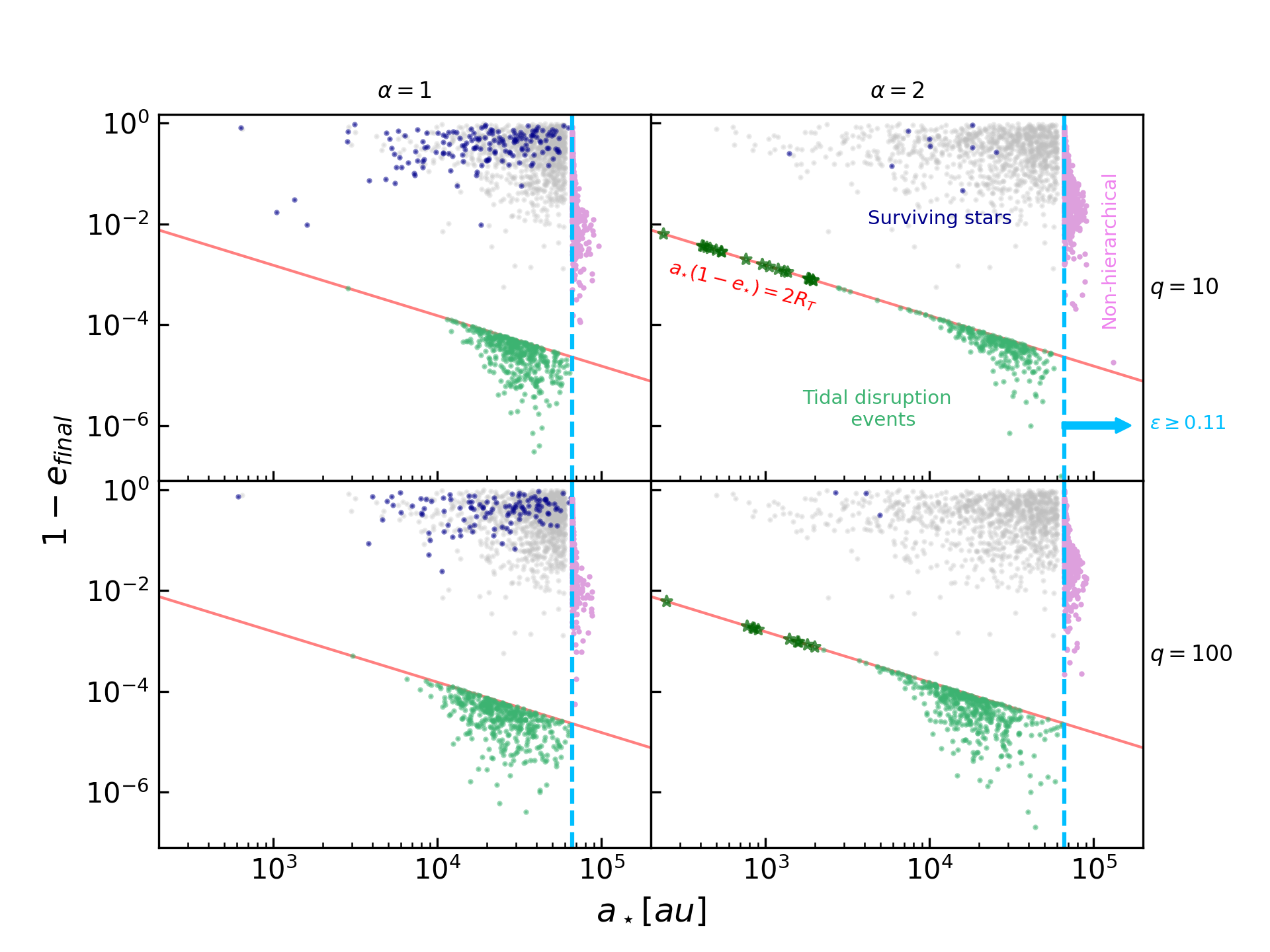}
  \end{center} 
  \caption{  \upshape {\bf Results of the Monte Carlo runs for $e_{\rm bin}=0.5.$} {We show a collection of systems that have undergone two-body relaxation and EKL. In grey, we indicate the initial conditions for the inner binary. Each panel has distinct initial conditions: the top panels have a BH mass ratio of 10, while the bottom panels have a BH mass ratio of 100; left panels have $\alpha=1$ while right panels have $\alpha=2$. Systems that end near their starting point are marked in dark blue, while those to the right of the dashed light-blue line have crossed the threshold for maintaining their hierarchical triple system configuration and thus are marked as non-hierarchical in pink. Systems below the red line, resulting in the crossing of the tidal radius, are marked as TDEs in green. Green stars on the $R_{T}$ line are considered to be repeated TDEs on orbital periods $P_\star \leq 30$~yrs. As shown, more systems become tidally disrupted or non-hierarchical for an $\alpha$ of 2.}} \label{fig:Scatter} 
\end{figure*}


We chose two representative masses of the companion of $m_2=10^8$~M$_\odot$ and $10^9$~M$_\odot$. We explore three outer orbit eccentricities of $e_{\rm bin} = 0.3,0.5,0.7$. The semi-major axis of the companion was set to be half the sphere of influence of the primary for all runs.

For the stellar population, we chose two representative density profiles, $\alpha = 1$ and $2$, following Equation \ref{eq:rhostar}. Furthermore, we adopted a thermal eccentricity distribution for the stars while the argument of periapsis and longitude of ascending nodes were chosen from a uniform distribution between $0-2 \pi$. The mutual inclination was chosen from an isotropic distribution, i.e., uniform in $\cos i$. 
See Table \ref{table:simparams} for the list of parameters chosen for the 12 realizations.

Figure \ref{fig:Scatter} shows a representative example of the runs, where the grey points mark the initial conditions. In total, we see four distinct outcomes  (the color code in Figure \ref{fig:Scatter}, as well as \ref{fig:Scatter0.3} and \ref{fig:Scatter0.7} from the Appendix, is mentioned below): 
\begin{enumerate}
    \item During the evolution the star's pericenter crossed $R_T$, see Eq.~(\ref{eq:Rcrit}).  We mark these systems as stars that become tidally disrupted. These systems are marked in green points below the $a_\star (1-e_\star) = R_T$ line. The fraction of systems that we mark TDEs is shown as the upper limit in the 6th column ($f_{\rm TDE}$) of Table \ref{table:simparams}. The lower limit of the TDE fraction includes only TDEs that are within the Hill radius of the primary SMBH. 
    \item Systems that have crossed the $R_T$ threshold and have periods shorter or equal to $30$~yrs we mark as repeated TDEs ($f_{\rm rTDE}$). These systems are highlighted as green stars.
    \item System evolved up to $1$~Gyr, without crossing the $R_T$, we consider as {\it survived}. These systems are marked in dark blue. 
    \item Finally, in some cases, the system violated the hierarchical condition, Eq.~(\ref{eq:epsilon}). These systems are marked as pink and reside to the right of the vertical hierarchical line. The fate of these systems is somewhat unclear, but as discussed in \citet{naoz_combined_2022}, we expect these stars to have their eccentricity excited to large values \citep[e.g.,][]{Bhaskar+21}, and thus, perhaps all of them will end as TDEs. We consider this case as an upper limit for the TDE fraction  (see Table \ref{table:simparams}, 7th column ($f_{\rm TDE}$ $+$ $ f_{\rm non-hierarchical}$)). 
\end{enumerate}



The evolution of systems with a stellar density distribution of $\alpha = 1$ (leftmost panels of Figure \ref{fig:Scatter}) showcase stars ending in the three main outcomes: stars surviving, stars ending beyond the hierarchical edge, and stars becoming TDEs. Systems with $\alpha = 2$ (rightmost panels of Figure \ref{fig:Scatter}) experience most systems having either the stellar component leaving the hierarchical edge  or stellar components becoming tidally disrupted.

Not surprisingly, fewer systems survive when $\alpha=2$, since this density profile results in a shorter timescale, i.e., larger change to the angular momentum $\Delta h/h$, as shown in Figure \ref{fig:timescale}. See \citet{naoz_combined_2022} for a detailed discussion. For the $\alpha=1$ case, more systems survive since this density profile yields a longer timescale, which results in a smaller change to the angular momentum, as depicted in Figure \ref{fig:timescale}. However, compared to the EKL-only scenario (\citet{Mockler+22} in prep. results), we still find the inclusion of two-body relaxation enhances the rate of systems that result in TDEs. 

\begin{figure}
  \begin{center} 
    \includegraphics[width=\linewidth]{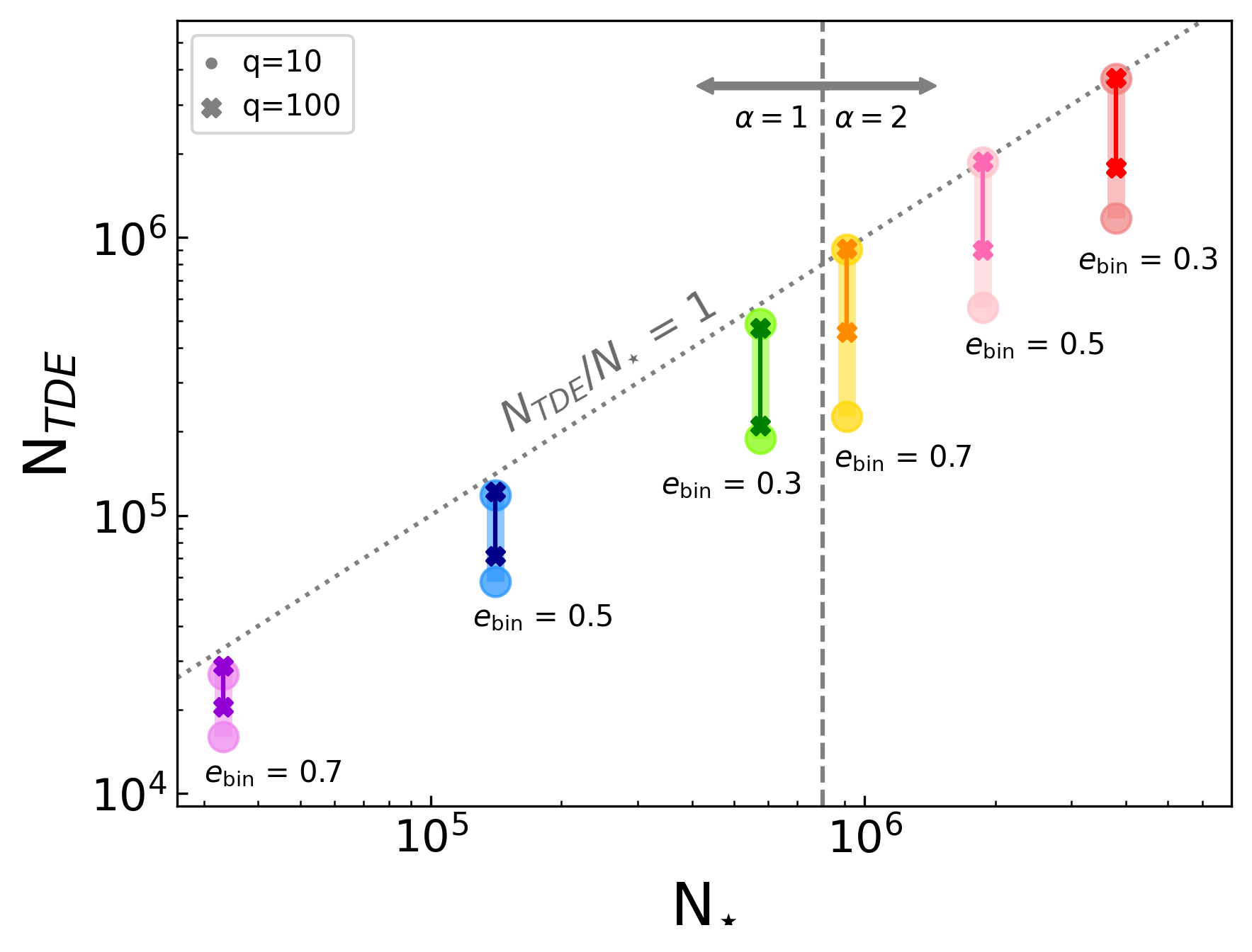}
  \end{center} 
  \caption{  \upshape {\bf Comparing the number of stars to the number of tidally disrupted events.} {Here, the eccentricity of the outer binary is varied on values $e_{\rm bin} = 0.3, 0.5, 0.7$. Mass ratios of $q = 10$ and $q = 100$ are also compared and are indicated by circles and ``$\boldsymbol{\times}$'s," respectively. Pairs of  $e_{\rm bin}$ and $q$ are indicated by different colors, as seen in the legend. The stellar density profile is changed, and a dashed line shows $\alpha = 1$ values to the left and $\alpha = 2$ to the right. A linear trend is seen in the number of TDEs as $e_{\rm bin}$ decreases and $\alpha$ increases.}  } \label{fig:numTDEs} 
\end{figure}


\section{Predictions and observational signatures}\label{sec:predictions}
\subsection{TDE rate}\label{sec:TDErate}
A key factor in estimating the TDE rate is the number of stars in the sphere inside the hierarchical edge $N_{\star}(r\leq r_{\rm max})$. This number is sensitive to the density profile as well as to the eccentricity of the SMBH binary. To estimate the number of stars within this range, $N_{\star}(\leq r_{\rm max})$, we use the $M-\sigma$ relation:
\begin{equation}\label{eq:star}
    N_{\star}(\leq r_{\rm max}) =  \frac{ m_1}{\langle m_\star \rangle}\left(\frac{G\sqrt{m_1 M_0}}{\sigma_0^2 r_{\star}}\right)^{-3+\alpha} \nonumber \ ,
\end{equation}
and 
\begin{equation}
    r_{\rm max} = \frac{a_\star}{\epsilon} e_{\rm bin} (1-e_{\rm bin}^2) \ ,
\end{equation} where we take $\epsilon=0.1$. Given this number of stars and the fraction of systems that become TDEs for all simulation runs (see Table \ref{table:simparams}), we can estimate the number of stars that become TDEs, $N_{\rm TDE}$. Figure \ref{fig:numTDEs}, depicts $N_{\rm TDE}$ as a function of $N_{\star}(\leq r_{\rm max})$, for the different density profile values (see arrows), eccentricity, and mass ratios. As shown in Figure \ref{fig:numTDEs}, the cusp profile, $\alpha=2$, which has a larger constriction of stars closer to the SMBH, compared to $\alpha=1$, results in a larger number of stars. 

We calculate the rate of TDEs for the different values of the power law of the stellar distribution ($\alpha$), the outer orbit eccentricity ($e_{\rm bin}$), and the mass ratio ($q$), as outlined in Table \ref{table:simparams}.  The calculated TDE rates from our simulations are then compared to the average observed TDE rate as taken from \citep[][]{van_velzen_optical-ultraviolet_2020} and the observed post-starburst (PSB) TDE rate from \citep[][]{french_structure_2020}, shaded regions in Figure \ref{fig:rate_alpha}. The calculated rates span from a minimum which is chosen by including all the hierarchical systems that end up as TDEs (indicated as green dots and green stars below the solid critical line in Figure \ref{fig:Scatter}). The maximum rate additionally includes all the non-hierarchical systems (indicated as the light pink dots to the right of the dashed lines in Figure \ref{fig:Scatter}). These non-hierarchical systems are likely to be excited to high eccentricities \citep[e.g.,][]{Bhaskar+21} and may eventually
become TDEs \citep[similarly to the maximum rate of EMRIs][]{naoz_combined_2022}.

We point out that some of the stars reside beyond the Hill radius of the SMBH binaries. However, as was shown by \citep[e.g.,][]{Zhang+23}, this configuration does not produce an instantaneous destabilization of the star's orbit. Rather the system can undergo eccentricity excitations before changing the energy of the stellar orbit (i.e., the semi-major axis). We find that in most of the systems, the destabilization timescale \citep[following][]{Zhang+23} is larger than the time to become a TDE. We, thus, adopt the lower limit of the rate to be the TDEs within the hierarchical limit (set by $\epsilon \leq 0.1$, see Eq.~(\ref{eq:epsilon})).     

Consider first the comparison between the calculated rates as a function of $\alpha$. As shown above, $\alpha$ is one of the key parameters that affect the combined effect of EKL with 2-body relaxation (see Figure \ref{fig:timescale}).  Thus, in Figure \ref{fig:rate_alpha}, we compare the results between $\alpha=1$ and $\alpha=2$, left and right column, respectively, for the simulations with  $e_{\rm bin}=0.3,0.5$, and $0.7$ (see labels), and for $q=10$ and $q=100$, top and bottom panels, respectively.
As depicted, the $\alpha=2$ case yields a slightly higher rate, with somewhat less dependency on the eccentricity of the SMBH binary eccentricity $e_{\rm bin}$. The $\alpha=2$ case represents a cusp density profile for which the two-body relaxation angular momentum changes are significant over a larger part of the parameter space. Thus, this behavior assists the EKL eccentricity excitations nearly independent of the binary eccentricity. 

However, several studies suggested that the stellar distribution in our own Milky Way's galactic center is core-like rather than cusp, with $\alpha$ closer to unity \citep[e.g.,][]{Lu+19,Schodel+17,Schodel+18,Schodel+20,Gallego+20}. Thus, if our galactic center is representative of galaxy nuclei's stellar distribution, the left column of Figure \ref{fig:rate_alpha} may be a more representative scenario. In this case, the weak kicks due to two-body relaxation are important but do not wash out the EKL sensitivity to the outer orbit eccentricity. Thus, as depicted in this Figure, lower SMBH eccentricity yields a higher TDE rate (simply more stars). We see that the $e_{\rm bin}=0.5$ TDE rate is consistent with the post-star burst TDE observed rate. Thus this may highlight a preferred SMBH formation channel \citep[e.g.,][]{dotti_massive_2012}. 

We remind the reader that the SMBH binary was set at half the distance to the sphere of influence. The latter dictates the number of stars, within that sphere, via the $M-\sigma$ relation. Thus, setting the number of stars that can potentially  undergo TDEs. Therefore the dependency on the eccentricity is degenerate with the dependency of the SMBH binary separation, which is beyond the scope of this paper.

\begin{figure}
  \begin{center} 
    \includegraphics[width=\linewidth]{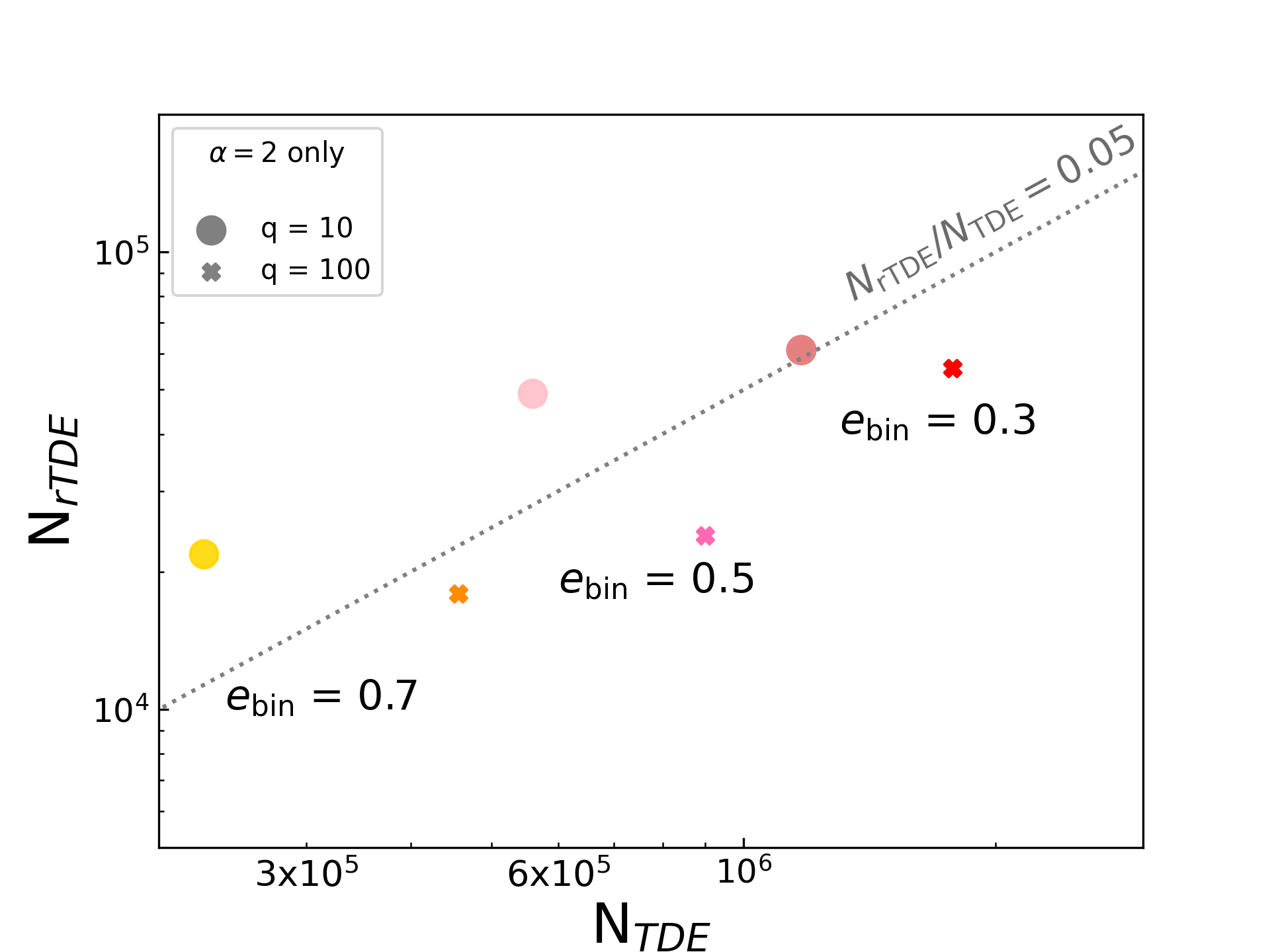}
  \end{center} 
  \caption{  \upshape {\bf Comparing the number of repeating tidally disrupted events to the total number of tidal disruption events.} {The eccentricity of the outer binary is varied on values $e_{\rm bin} = 0.3, 0.5, 0.7$. Mass ratios of $q = 10$ and $q = 100$ are represented as circles and ``$\boldsymbol{\times}$'s," respectively. The colors represent the pairs of $e_{\rm bin}$ and $\alpha$. Here, unlike in Fig. \ref{fig:numTDEs}, only $\alpha = 2$ is shown as only this cuspy profile produced rTDEs as seen in Fig.\ref{fig:Scatter}, \ref{fig:Scatter0.3}, \ref{fig:Scatter0.7}.} } \label{fig:numrTDEs} 
\end{figure}

\subsection{Formation of repeated TDEs}\label{sec:rTDE}
rTDE observations present many intriguing puzzles. For example, how do these stars get only partially disrupted? This question was addressed in earlier studies \citep[][]{guillochon_hydrodynamical_2013,coughlin_variability_2015} and is still under investigation. Here we focus on the question of how these stars get to such short separations from the SMBH without having their eccentricity excited  to high values at larger distances from the SMBH. 

The combined effect of EKL with two-body relaxation slightly changes the star's SMA due to kicks. Suppose the star migrates to a regime where EKL is more efficient (for example, a higher $\epsilon$ or an inclination closer to $90^\circ$). In that case, the star will undergo large eccentricity excitations due to EKL. This behavior is depicted in Figure \ref{fig:timeEvolution}, where the star undergoes faster EKL oscillations, and its eccentricity is excited to higher values as its SMA slightly increases. Furthermore, since the kicks are proportional to the star's orbital velocity (Eq.~(\ref{eq:Deltav})), the kick can be larger for a star initially closer (larger velocity). The example system shown in Figure \ref{fig:timeEvolution} highlights this behavior, and we outline this signature below. 

We arbitrarily choose a period of $30$~yrs to mark the rTDEs candidates. As for the non-repeating TDEs, the number of expected rTDEs is proportional to the number of stars in the sphere of influence (see Figure \ref{fig:numrTDEs}).  As depicted in Figure \ref{fig:Scatter}, the initial cuspier density distribution (i.e., $\alpha=2$) results in a larger fraction of rTDEs.  
Thus, these results suggest that detections of rTDEs may be used as an indicator for the stars underlying density profile. 

\begin{figure*}
  \begin{center} 
    \includegraphics[width=\linewidth]{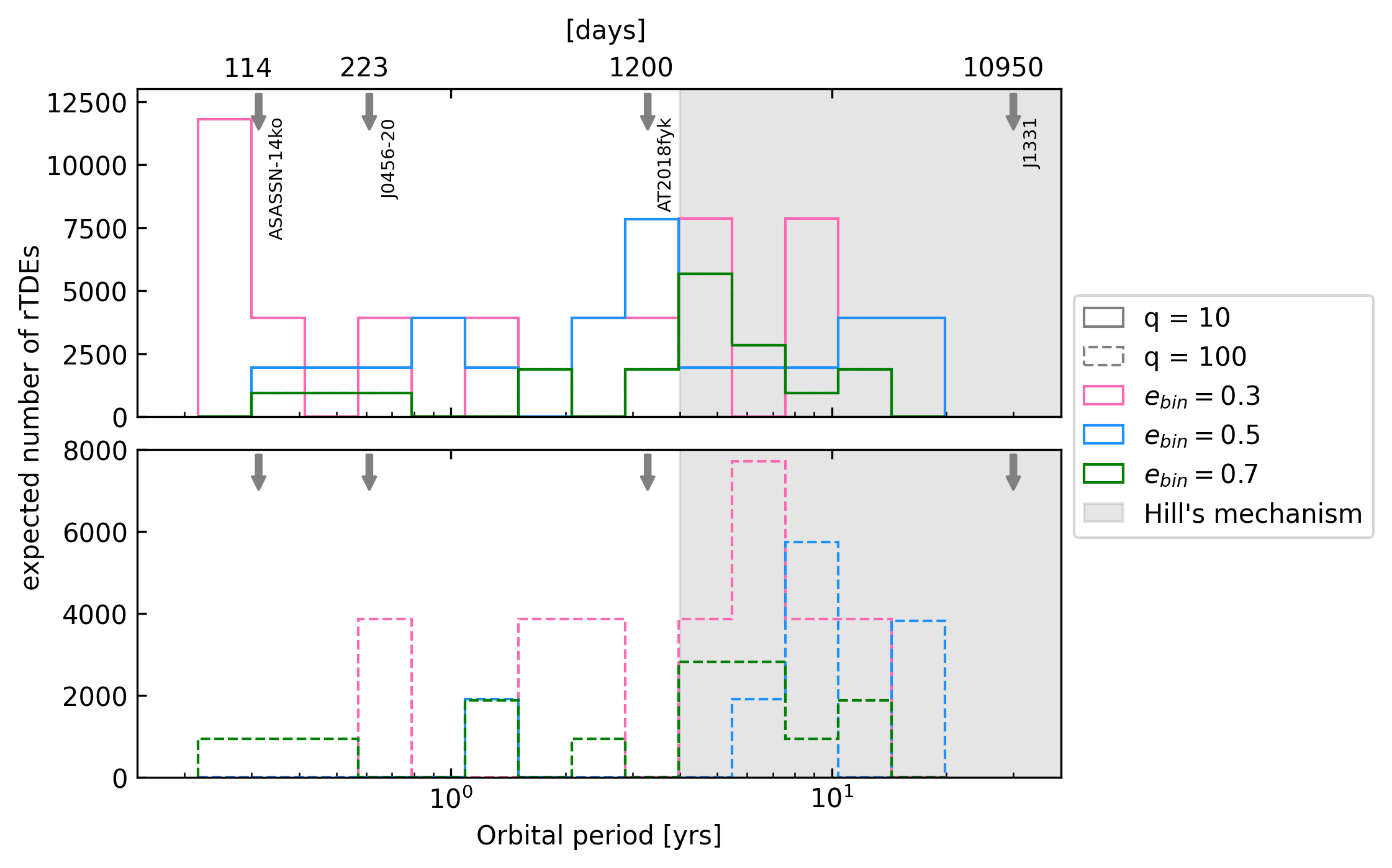}
  \end{center} 
  \caption{  \upshape {\textbf{Period distribution of expected rTDEs.} The expected number of rTDEs, normalized by the $M-\sigma$ relation, is shown for different pairs of binary eccentricity and mass ratio, top panel q = 10 and bottom panel q = 100. Arrows show the current observations of potential rTDEs. The shaded grey region spans the parameter space for which the Hill's mechanism produces rTDEs (see text for details).}} \label{fig:rTDE_period} 
\end{figure*}

Figure \ref{fig:rTDE_period} shows the expected rTDE period distribution adopting the M-$\sigma$ normalization explained above. As shown, this mechanism results in a period distribution consistent with known observations of potential repeated TDEs with periods ranging hundreds of days up to $30$ years, \citep[e.g.][]{payne_rapid_2022,wevers_live_2023,liu_deciphering_2023,malyali_rebrightening_2023}. Note that our simulations do not suggest a preferred initial semi-major axis that rTDEs originate from in the star cluster based on this channel. It was suggested that the Hill's mechanism may contribute to the formation of rTDEs as recently proposed by \citet{cufari_using_2022}. However, one can relate the period of the captured star from the Hill's mechanism as a function of the separation of the inner binary \cite[e.g.,][]{hills_possible_1975,yu_ejection_2003}.  A binary undergoes many weak gravitational interactions with neighboring stars during its lifetime. These encounters tend to widen and unbind the binary \citep[e.g.,][]{binney_galactic_2008}.  Thus, in order for a binary to remain bound before the Hill's mechanism takes place, its semi-major axis should be smaller than a critical value and larger than at least two times the radius of the binary members \citep{Rose+20}.  Taking the minimum separation between a binary (i.e., assuming an extremely hard binary that does not evolve), we can find the corresponding minimum period for a repeated TDE.  
We show this period in Figure \ref{fig:rTDE_period}, suggesting the long period J1331 could have originated from the Hills mechanism, while the other potential rTDEs seem to be consistent with the channel described here.





\subsection{Comparison with TDE without two-body relaxation}

Combined with EKL, two-body relaxation efficiently drives the stars onto the SMBH. As discussed above, the efficiency depends on the number of stars within the hierarchical limit (thus the SMBH binary separation of their eccentricity) and the density profile. The latter is largely unknown within the inner parsec of observed galaxies. Currently, some of the best estimations can reach as close as $30$~pc to a few kiloparsecs  \citep[e.g.,][]{french_structure_2020}, finding a cusp-like distribution. On the other hand, the stellar distribution of our own galactic center seems to follow a core-like distribution \citep[e.g.,][]{Genzel+03,gallego-cano_distribution_2018} 


In  Figure \ref{fig:rate_compare}, we compare the combined effect with an EKL-only approach adopted from \citet{Mockler+22}. The Figure shows the time-dependent TDE rate for two examples of $m_1 = 10^7$~M$_\odot$, $m_2 = 10^8$~M$_\odot$, $m_\star = 0.8$~M$_\odot$, $e_{\rm bin} = 0.5$, $a_{\rm bin} \approx 2$~pc with the violet shade indicating the system with $\alpha = 1$ and the cyan shade $\alpha = 2$.
We note that in the case of EKL-only, there is no system that is pushed beyond the hierarchical limit (systems beyond the vertical line in Figure \ref{fig:Scatter}). Thus, the maximum value calculated differs between the two runs. However, as can be seen, the possible additional non-hierarchical TDEs do not yield a significant difference, especially for the shallow, $\alpha = 1$ profile. It is worth noting that the combined effect results in a longer, extended time-dependent rate, perhaps allowing star formation to occur and replenish the TDEs. 

As expected, the EKL-only channel produces a burst-like event, depicted as a shorter rate in Figure \ref{fig:rate_compare}. This behavior was noted in \citet{Naoz+14} and \citet{Naoz+22}. In particular, we suggest that systems with a core-like density distribution, thus a longer two-body relaxation timescale, may undergo a burst-like rate. In other words, these systems may have already formed their TDEs and, without newly formed stars, are less likely to produce a TDE at a high rate. 

 
\subsection{Combined TDE and EMRI Events}

The rates suggested here \cite[and in ][]{Mockler+22} are much higher than previously considered when considering single SMBHs \citep[e.g.,][]{stone_rates_2016}.
Given the right conditions, in terms of density profile and the SMBH binary separation and eccentricity, Figures \ref{fig:rate_alpha} and \ref{fig:numTDEs} suggest that a galaxy may have multiple TDEs. Similarly, for the right conditions, a galaxy may have repeated TDEs (as suggested by Figures \ref{fig:Scatter} and \ref{fig:numrTDEs}). Beyond TDEs, it was recently suggested that the similar physical processes described here could yield much higher extreme mass ratio inspiral (EMRIs) rates \citep[orders of magnitude than estimated before][]{Naoz+22}. Thus, by combing the two findings, we speculate that a combined TDE and EMRI event may have a non-negligible rate. 




\section{Discussion}\label{sec:discussion}


\begin{figure*}
  \begin{center} 
    \includegraphics[width=\linewidth]{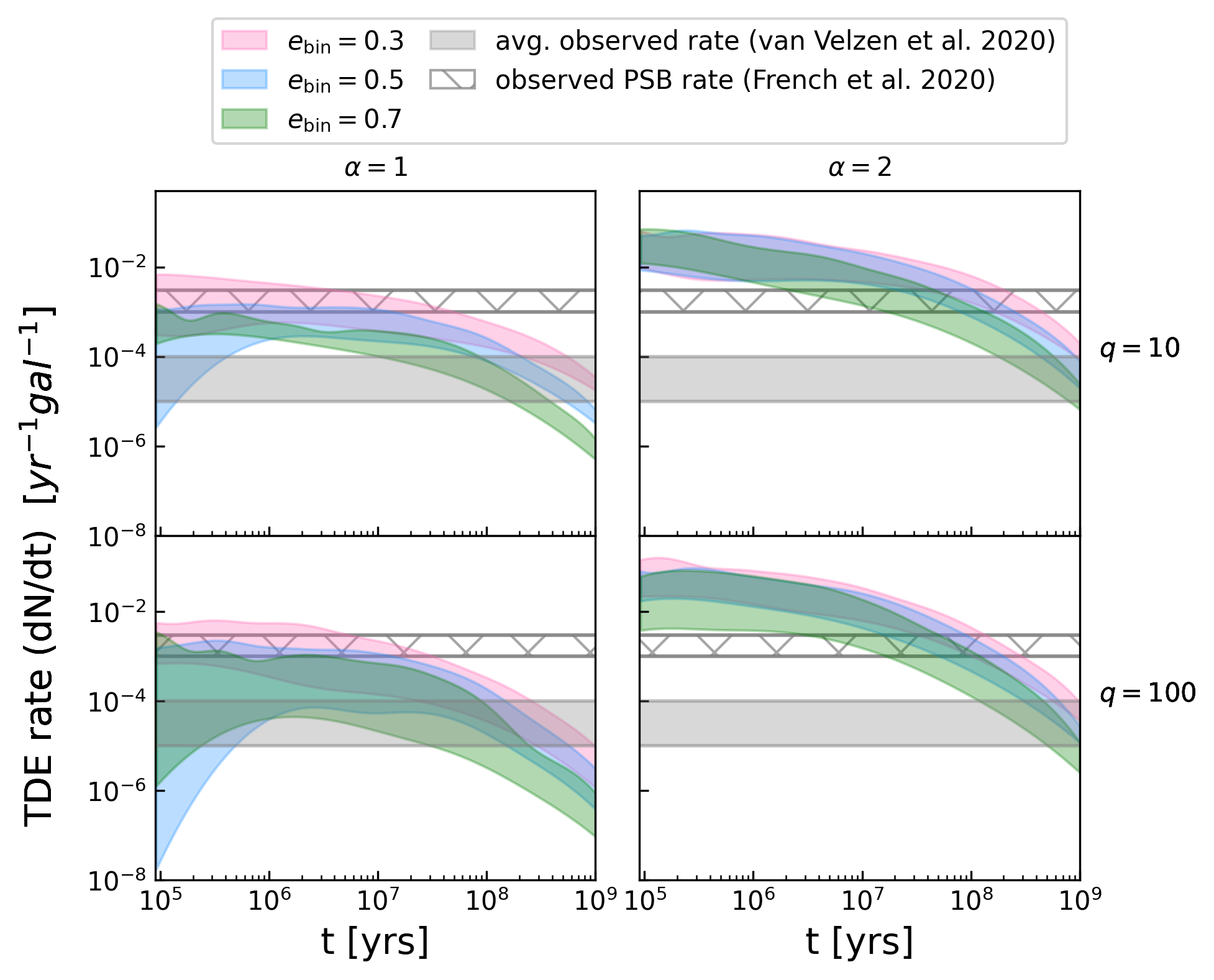}
  \end{center} 
  \caption{  \upshape {\bf Calculated TDE rates.} The rate for all systems is calculated and compared to both the average observed rate and the observed PSB rate. The left columns show systems with the density distribution power-law $\alpha = 1$ while the right column panels show that of $\alpha = 2$. The top panels have a mass ratio of $q = 10$, and the bottom panels $q = 100.$ As seen, $\alpha = 2$ systems produce TDEs at an enhanced rate with little dependence on eccentricity.   } \label{fig:rate_alpha} 
\end{figure*}

\begin{figure}
  \begin{center} 
    \includegraphics[width=\linewidth]{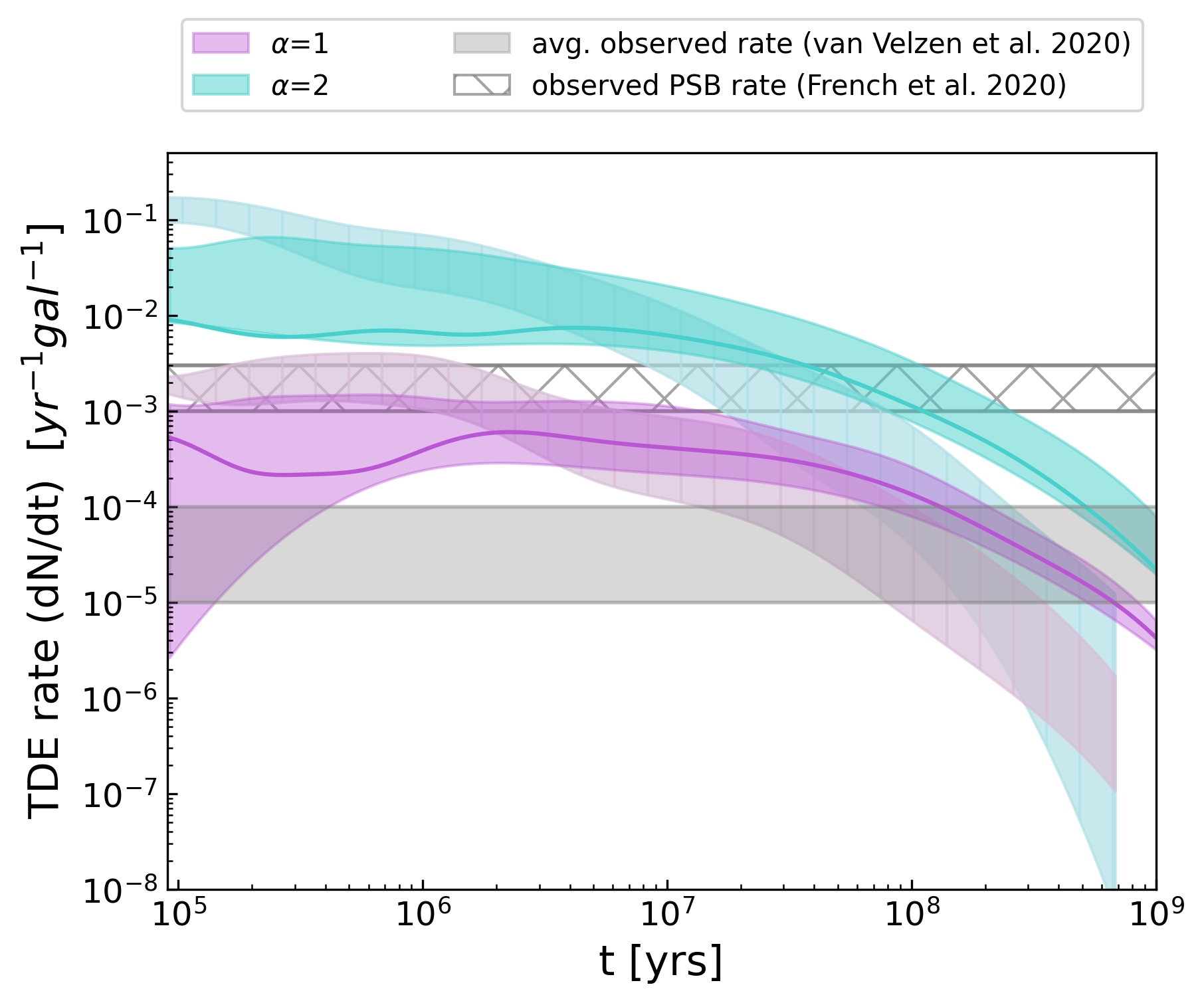}
  \end{center} 
  \caption{  \upshape {\bf TDE rates with and without two-body relaxation.} Rates shown here are for systems with two distinct stellar distributions $\alpha = 1$ in violet and $2$ in cyan. The shaded rates are for systems that underwent both two-body relaxation and EKL (+GR), while the hatched grey-like rates are for those with only EKL (+GR) processes as adapted from \citet{Mockler+22}. The adapted rates showcase high peaks on shorter timescales, while rates in this work have lower peaks and are extended further in time. The minimum for all rates is calculated  only for our conservative lower limit for TDEs, which are constrained within the Hill radius. The maximum for the EKL-only rates includes the systems that are beyond the Hill and the Roche limits. This limit corresponds to the solid line in the Figure. The maximum value for the rates for the EKL (+GR) + 2 body Relaxation includes the systems beyond the hierarchical limit as well. 
  } \label{fig:rate_compare} 
\end{figure}

We demonstrate that the combined physical processes of two-body relaxation and EKL in SMBH binaries significantly affect the dynamics of TDEs.
Two-body relaxation has been proposed as one of the most promising physical processes to form TDEs efficiently. In this process, weak two-body kicks from the population of stars surrounding the SMBH can change the star's orbit over time, plunging it into the SMBH from large distances. Perturbations from an SMBH companion via the EKL mechanism can also excite the star to high eccentricities, providing another channel for forming TDEs.  

Here we demonstrated that the two body relaxation, combined with the EKL-induced eccentricity, plays a crucial role in forming TDEs, and rTDEs. In this case, stars experience high eccentricities due to the two-body relaxation process, which drives the stars into a more EKL-sensitive regime 
as demonstrated in Figure \ref{fig:timescale}. This combination not only leads to more TDEs than EKL on its own, but it also naturally produces repeating TDEs. Additionally, two-body relaxation leads to more stars scattering beyond the hierarchical radius (Figure \ref{fig:Scatter}, where we expect many to be disrupted \citep[e.g.,][]{Grishin+18,Bhaskar+21}.



In addition to combining the aforementioned two effects, we have also explored the effect of the star's density profile by comparing two extreme cases. One of a shallow profile, corresponding to $\alpha=1$, and the second of a cusp profile of $\alpha=2$ (see Figure \ref{fig:Scatter}). Observations of the galactic center suggest that the stellar distribution in the inner $0.1$~pc may be shallow, i.e., closer to $\alpha=1$ \citep[e.g.,][]{Genzel+03,Schodel+17,Schodel+18,Schodel+20}. While it may be that other galactic nuclei have similar conditions to our Galactic center, TDEs have been preferentially found in post-starburst galaxies, which may have different properties \citep[e.g.,][]{yang_detailed_2008,french_tidal_2016, law-smith_tidal_2017,2021ApJ...907L..21D}; thus, we also considered a cusp profile (i.e., $\alpha=2$). 
A shallower density profile has two main consequences. First, it creates a longer two-body relaxation timescale, thus, the angular momentum changes via the 2-body scattering are smaller compared to the EKL (see bottom panel of Figure \ref{fig:timescale}). Second, it effectively changes the number of stars inside the hierarchical limit, which has a direct consequence on the number of stars resulting in TDEs and rTDEs (as highlighted in Figures \ref{fig:numTDEs} and \ref{fig:numrTDEs}).

We also investigated the effect of the SMBH binary's mass ratio and eccentricity. We show representative results of our Monte Carlo simulations in Figure \ref{fig:Scatter} (see Figures \ref{fig:Scatter0.3} and \ref{fig:Scatter0.7} in the Appendix for the configurations involving $e_{\rm bin}=0.3$ and $e_{\rm bin}=0.7$).
As seen in Figure \ref{fig:rate_alpha}, the rate is nearly independent of the binary eccentricity. However,  $e_{\rm bin} = 0.3$ tended toward a higher and more extended rate compared to the higher eccentricities. This is a direct result of the fact that lower eccentricity SMBH binary configuration allows for more stars to reside in the hierarchical limit (see for reference Figure \ref{fig:numTDEs}). Thus, as shown in \ref{fig:rate_alpha}, the rate is largely insensitive to the mass ratio and the orbital eccentricity of the SMBH binary\footnote{Note that as depicted in Figure \ref{fig:rate_alpha} the SMBH eccentricity yields a slight difference in the rate for a core-like distribution. This behavior is expected because the two-body relaxation is somewhat less dominant compared to the EKL exaction, and the latter is susceptible to the eccentricity \citep[e.g.,][]{li_chaos_2014,li_eccentricity_2014}. }.

We note that we have neglected possible stellar collisions or star-compact object collisions, which may have significant consequences on the masses of stars in galactic nuclei \citep[e.g.,][]{Rose+22,rose+23}. Additionally, a substantial fraction of stars is expected to be in binaries (see \citet{Raghavan+10}, even in the galactic center \citet{Naoz+18}). Collisions, two-body relaxation, and weak perturbations that may unbind the binary stars may also affect the orbital configuration of these binaries \citep[e.g.,][]{Rose+20}. The EKL mechanism can also merge binary stars together \citep[e.g.,][]{stephan_merging_2016,Stephan+19}. In general, we expect that binaries that undergo the combined effect of two-body relaxation and EKL from an SMBH companion will yield a possible Hills \citep[e.g.,][]{hills_possible_1975} process or, in some cases, even double TDE \citep[e.g.,][]{Mandel+15}. The details of such a process are beyond the scope of this paper.

 The main contributor to the TDE rate is the underlying stellar distribution and, of course, the existence of the SMBH binary. Therefore comparing to observations, this may help constrain the stellar cusp- or core-like distribution and the frequency of SMBH binaries. Note that increasing the cusp of the density profile yields a shorter two-body relaxation timescale compared to a core-like profile. While this was noted before for a single SMBH system, \citet{stone_delay_2018}, here we report an opposite dependency as a function of the SMBH (disruptor) mass and thus the number of stars. In other words, since the two-body relaxation timescale is increasing as a function of the SMBH mass, a single SMBH system predicts a decreasing rate as the SMBH mass. In the case of the combined two-body relaxation with EKL (for a given mass ratio), we find the opposite trend, i.e., an increasing rate with the disruptor mass (see Figure \ref{fig:numTDEs}). A similar result was obtained for the EMRIs rate as a function of the SMBH mass; see figure 5 in \citet{naoz_combined_2022}.

As shown by \citet{Mockler+22}, the EKL mechanism leads to a unique signature of TDEs on the smaller SMBH companion. Combining the two-body relaxation processes and the EKL mechanism leads to higher eccentricity excitations, yielding an extended time-dependent rate (see Figures \ref{fig:rate_alpha} and \ref{fig:rate_compare}). Additionally, we demonstrated that combining EKL and two-body relaxation can be used as a signature of the underlying cusp of the less massive SMBH in an SMBH binary configuration. Further, we suggested that this combined dynamical effect can naturally give rise to rTDEs. Lastly, these results can be used to constrain the SMBH binary fraction. In other words, we suggest that since these predictions only rely on one main assumption, i.e., the existence of binary SMBHs, they can be used to constrain SMBH binary fractions.

\begin{acknowledgments}
D.M. acknowledges the partial support from NSF graduate fellowship DGE-2034835, the Eugene V. Cota-Robles Fellowship, and the NASA ATP  80NSSC20K0505. B.M. is grateful for support from the U.C. Chancellor's Postdoctoral fellowship and from Swift (80NSSC21K1409). S.N. acknowledges the partial support from NASA ATP 80NSSC20K0505 and from NSF-AST 2206428 grant as well as thanks Howard and Astrid Preston for their generous support. S.R. thanks the Nina Byers Fellowship, the Charles E Young Fellowship, and the Michael A. Jura Memorial Graduate Award for support, as well as partial support from NASA ATP 80NSSC20K0505. E.R.R. thanks the Heising-Simons Foundation, NSF (AST-1615881 and AST-2206243),
Swift (80NSSC21K1409, 80NSSC19K1391) and Chandra (22-0142) for support.

\end{acknowledgments}

%

\vspace{5mm}




\appendix
\section{The orbit's time evolution}\label{App:timeEvol}
\begin{figure*}
  \begin{center} 
    \includegraphics[scale=0.7]{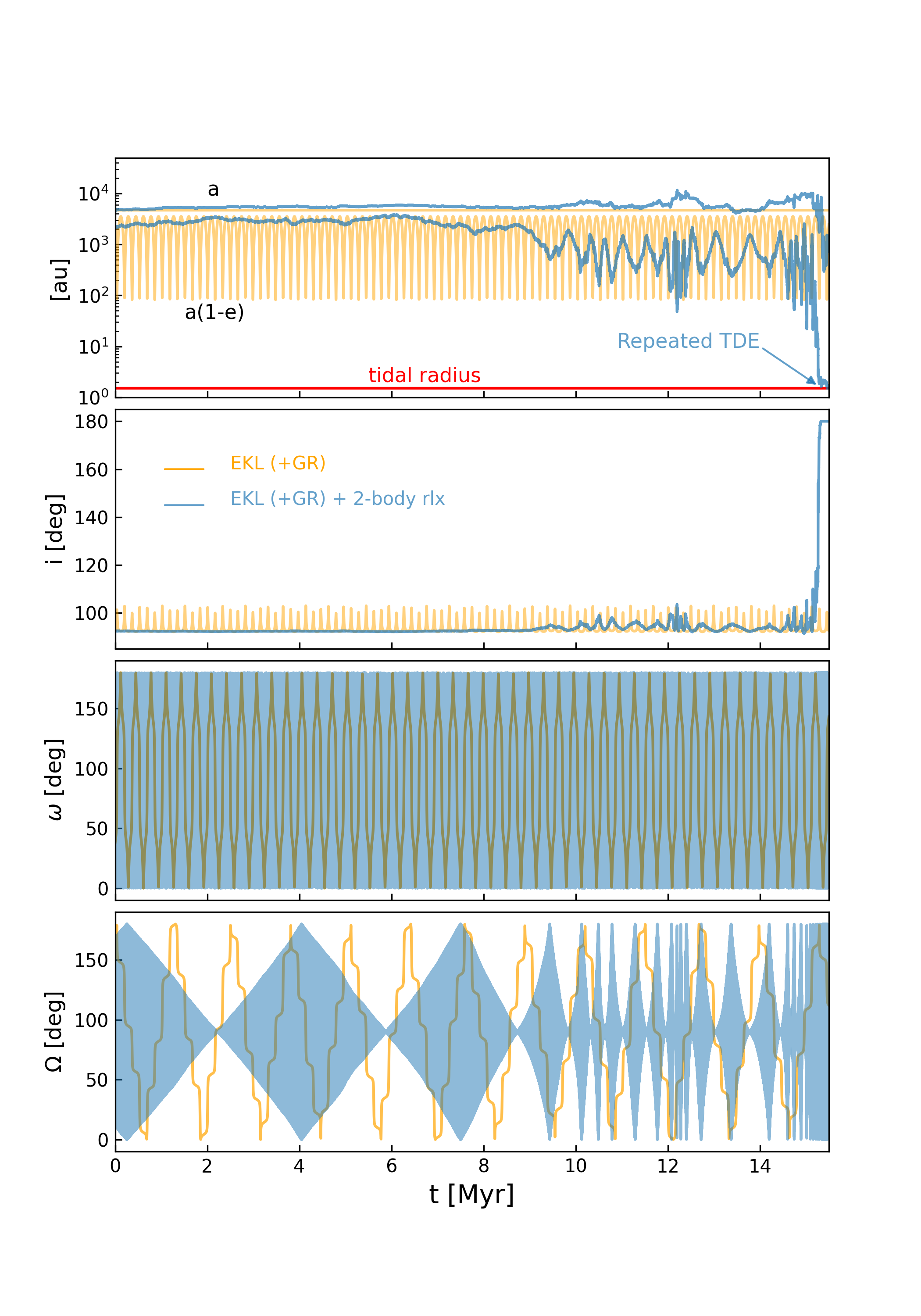}
  \end{center} 
  \caption{  \upshape {\bf Repeated TDE evolution including $\Omega$ and $\omega$.} Additional orbital elements of the same system in Figure \ref{fig:timeEvolution} illustrate the evolution of the argument of periapsis $\omega$, third panel, and the longitude of ascending nodes $\Omega$, bottom panel, both of which are taken from a uniform distribution between 0 and 2$\pi$. We compare the evolution of these orbital parameters as they undergo EKL (+GR) (yellow) to the combined effects of EKL (+GR) and two-body relaxation (blue). } \label{fig:evolution_omegas} 
\end{figure*}

The EKL mechanism is a resonant system at which the argument of periapsis ($\omega$) and the longitude of ascending nodes ($\Omega$) can undergo libration or rotation \citep[e.g.,][]{Li+14,Hansen+20}. The diffusive nature of the two-body relaxation can drive the system from libration to rotation \citep[as shown in][]{naoz_combined_2022}. In Figure \ref{fig:evolution_omegas} we illustrate the evolution of $\omega$ and $\Omega$, for both EKL (+GR) and the combined EKL(+GR) and two-body relaxation effects. While the evolution of $\omega$ and $\Omega$ appear solid, they oscillate rapidly for the EKL (+GR) and two-body relaxation case. We remind the reader that this system for the combined EKL(+GR) and two-body relaxation case is a repeated TDE system. 


\section{Additional Monte Carlo results}\label{App:MCresults}

\begin{figure*}
  \begin{center} 
    \includegraphics[width=\linewidth]{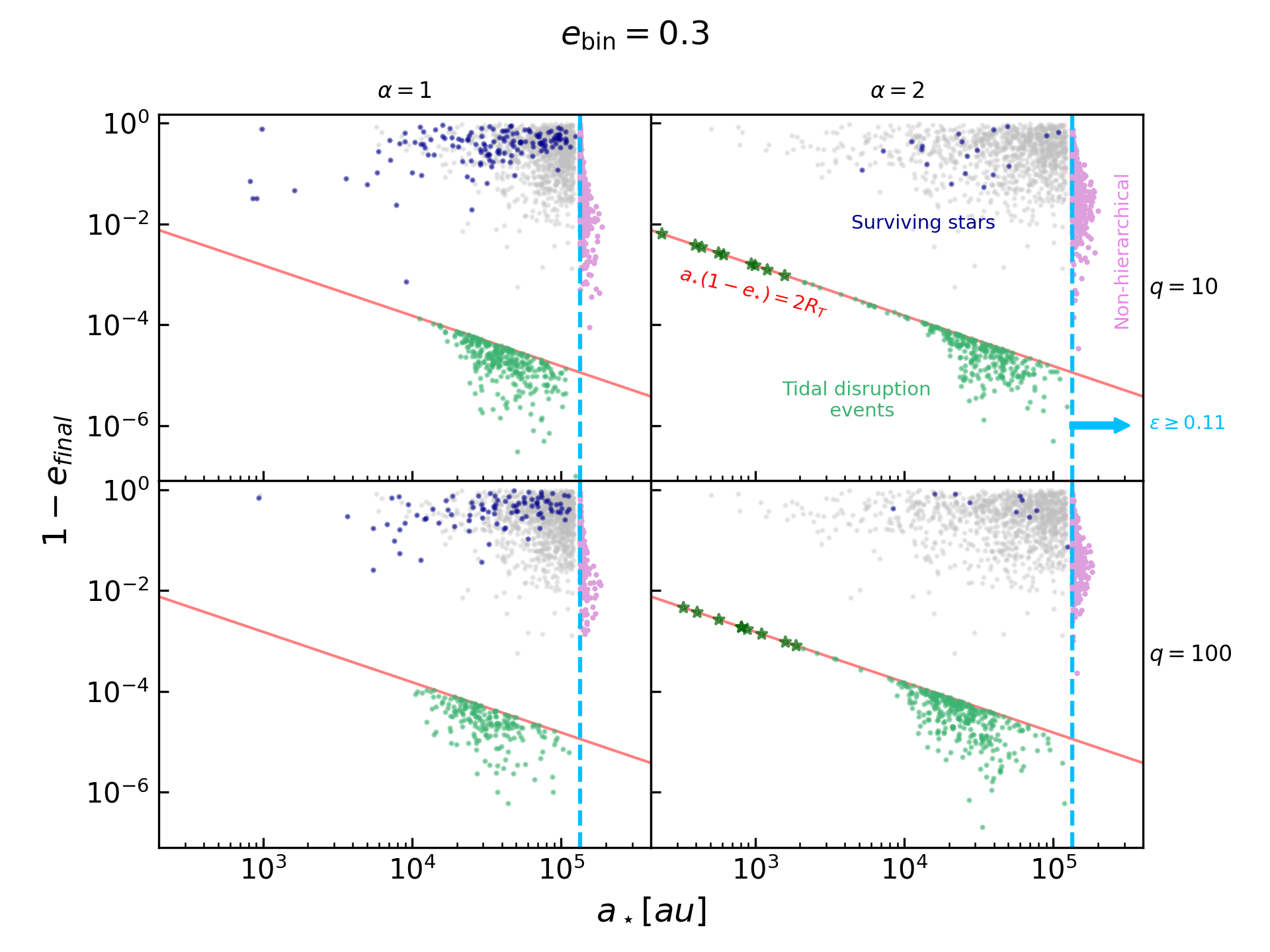}
  \end{center} 
  \caption{  \upshape {\bf Results of the Monte Carlo runs for e$_{\rm bin}$ = 0.3.} Similarly to Figure \ref{fig:Scatter}, we show a collection of systems that have undergone two-body relaxation and EKL, with the only change being the binary eccentricity reduced to e$_{\rm bin}$ = 0.3. As shown, only systems with $\alpha = 2$ produce rTDEs on orbital periods of 30 years or less. } \label{fig:Scatter0.3} 
\end{figure*}

\begin{figure*}
  \begin{center} 
    \includegraphics[width=\linewidth]{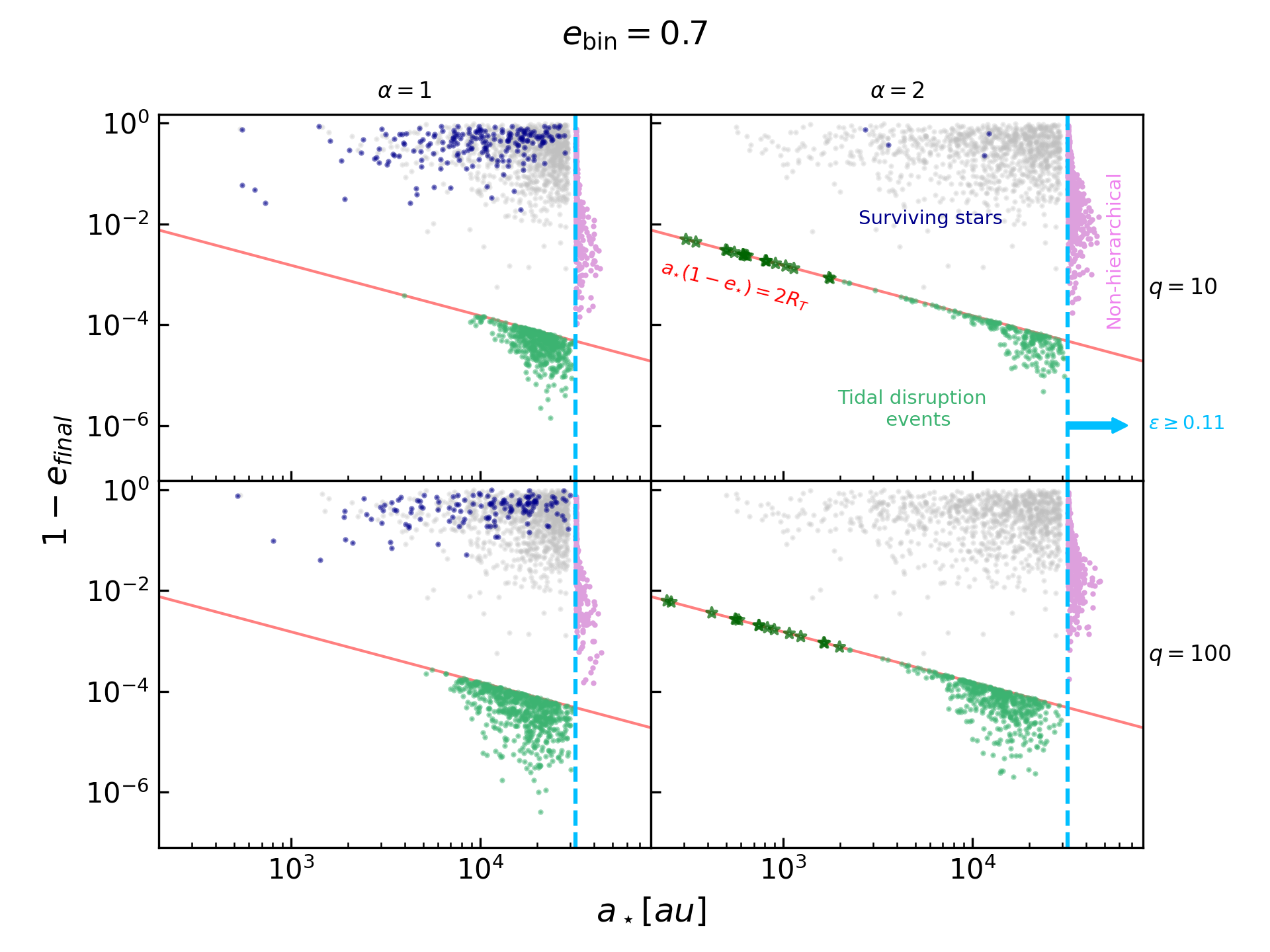} 
  \end{center} 
  \caption{  \upshape {\bf Results of the Monte Carlo runs for e$_{\rm bin}$ = 0.7.} Similarly to Figure \ref{fig:Scatter}, we show a collection of systems that have undergone two-body relaxation and EKL with the only change being the binary eccentricity increased to e$_{\rm bin}$ = 0.7. As shown, only systems with $\alpha = 2$ produce rTDEs on orbital periods of 30 years or less. } \label{fig:Scatter0.7} 
\end{figure*}

In Table \ref{table:simparams}, we described the full Monte Carlo simulations. As an example, in Figure \ref{fig:Scatter}, we showed the results from simulations with binary eccentricity $e_bin = 0.5$. Here we show additional systems undergoing two-body relaxation and EKL in Figure \ref{fig:Scatter0.3} and Figure \ref{fig:Scatter0.7}. Like in \ref{fig:Scatter}, these additional figures show the final evolution of several systems with different pairings of mass ratio, stellar density distribution, and binary eccentricity. In Figure \ref{fig:Scatter0.3}, the binary eccentricity is set to $e_{bin} = 0.3$, and in Fig. \ref{fig:Scatter0.7}, it is set to $e_{bin} = 0.7$. Similar to \ref{fig:Scatter}, repeated TDEs are only produced in the cuspier stellar density environments.  

\bibliographystyle{apj}
\bibliography{Binary, TDEs_in_SMBHB}
\end{document}